\documentclass[12pt,twoside]{article}
\usepackage{fleqn,espcrc1,epsfig}
\usepackage{graphicx}
\usepackage[figuresright]{rotating}

\newcommand{\AmS}{{\protect\the\textfont2
  A\kern-.1667em\lower.5ex\hbox{M}\kern-.125emS}}

\title{How can one  understand the lightest scalars, especially the $\sigma$}

\author{N.A. T\"ornqvist\thanks{Invited plenary talk by
N.A. T\"ornqvist at the ``Biennial Conference on Low Energy
Antiproton Physics'' (LEAP2000) in Venice, Italy, August 20-26,
2000. To appear in Nucl. Phys. A.} and A.D. Polosa
\address[MCSD]{Physics Department,
 POB 9, FIN-00014, University of Helsinki, Finland}%
        }

\begin{document}

% typeset front matter
\maketitle

\begin{abstract}
We discuss how the $a_0(980),\  f_0(980),\  K^*_0(1430)$ and
particularly the broad $\sigma$ resonance can be understood within
a coupled channel framework, which includes all light
two-pseudoscalar  thresholds together with constraints from Adler
zeroes, flavour symmetric couplings, unitarity and physically
acceptable analyticity. All $q\bar q$ scalars are, when
unitarized,  strongly distorted  by hadronic mass shifts, and the
nonstrange isoscalar state becomes a very broad resonance, with
its pole at 470-i250 MeV. We believe this is the sigma meson
required by models for spontaneous breaking of chiral symmetry.
Recently this light resonance has clearly been observed in
$D\to\sigma\pi\to3\pi$ by the E791 experiment at Fermilab, and we
discuss how this decay channel can be predicted in a Constituent
Quark Meson Model (CQM), which incorporates heavy quark and chiral
symmetries.

We also discuss the less well known phenomenon  that with a large
coupling there can appear  two physical resonance poles on the
second sheet although only one bare quark-antiquark state is put
in. The $f_0(980)$ and $f_0(1370)$ resonance poles can thus be two
manifestations of the same $s\bar s$ quark  state.  Both of these
states are seen clearly in $D_s\to 3\pi$  by the E791 experiment,
where $s\bar s$ intermediate states are expected to be dominant.
\end{abstract}
\def \gam {\frac{ N_f N_cg^2_{\pi q\bar q}}{8\pi} }
\def \gam {\frac{ N_f N_cg^2_{\pi q\bar q}}{8\pi} }
\def \gamm {N_f N_cg^2_{\pi q\bar q}/(8\pi) }
\def \be {\begin{equation}}
\def \ba {\begin{eqnarray}}
\def \ee {\end{equation}}
\def \ea {\end{eqnarray}}
\def \gap {{\rm gap}}
\def \gapp {{\rm \overline{gap}}}
\def \gappp {{\rm \overline{\overline{gap}}}}
\def \im {{\rm Im}}
\def \re {{\rm Re}}
\def \Tr {{\rm Tr}}
\def \P {$0^{-+}$}
\def \S {$0^{++}$}
\def \uu {$u\bar u$}
\def \dd {$d\bar d$}
\def \ss {$s\bar s$}
\def \qq {$q\bar q$}
\def \qqq {$qqq$}
\def \lsm {L$\sigma$M}
\def \sig {$\sigma$}
\def \gam {\frac{ N_f N_cg^2_{\pi q\bar q}}{8\pi} }
\def \gamm {N_f N_cg^2_{\pi q\bar q}/(8\pi) }
\def \be {\begin{equation}}
\def \ba {\begin{eqnarray}}
\def \ee {\end{equation}}
\def \ea {\end{eqnarray}}
\def\bea{\begin{eqnarray}}
\def\eea{\end{eqnarray}}
\def \gap {{\rm gap}}
\def \gapp {{\rm \overline{gap}}}
\def \gappp {{\rm \overline{\overline{gap}}}}
\def \im {{\rm Im}}
\def \re {{\rm Re}}
\def \Tr {{\rm Tr}}
\def \P {$0^{-+}$}
\def \S {$0^{++}$}
\def\zpp{$0^{++}$}
\def\fz{$f_0(980)$}
\def\az{$a_0(980)$}
\def\Kz{$K_0^*(1430)$}
\def\fzz{$f_0(1300)$}
\def\fzzz{$f_0(1200-1300)$}
\def\azz{$a_0(1450)$}
\def\ss{$ s\bar s $}
\def\uu{$u\bar u+d\bar d$}
\def\qq{$q\bar q$}
\def\KK{$K\bar K$}
\def\sig{$\sigma$}
\def\lsim{\;\raise0.3ex\hbox{$<$\kern-0.75em\raise-1.1ex\hbox{$\sim$}}\;}
\def\gsim{\raise0.3ex\hbox{$>$\kern-0.75em\raise-1.1ex\hbox{$\sim$}}}

\section{Introduction}
This talk is mainly based on earlier papers~\cite{NAT1,NAT2} on
the light scalars and on a more recent one \cite{gatto} on the
\sig\ in charm decay, including a few new comments. First we shall
discuss the evidence for the light \sig\ and explain how one can
understand the controversial light scalar mesons with a unitarized
quark model, which includes most well established theoretical
constraints:
\begin{itemize}
\item Adler zeroes as required by chiral symmetry,
\item all light two-pseudoscalar (PP) thresholds with flavor symmetric couplings in a coupled channel framework
\item physically acceptable analyticity, and
\item unitarity.
\end{itemize}  A unique
feature of this model is that it simultaneously describes the
whole scalar nonet and one obtains a good representation of a
large set of relevant data. Only six parameters, which all have a
clear physical interpretation, are needed.

After describing our understanding of the $q\bar q$ nonet, we
discuss the recently measured $D\to\sigma\pi\to 3\pi$ decay, where
the \sig\ is clearly seen as the dominant peak.
\section{The problematic scalars and the existence of the \sig }
The interpretation of the nature of lightest scalar mesons has
been controversial for long. There is no general agreement on
where are the  $q\bar q$ states, is there a glueball among the
light scalars, are some of the scalars multiquark or $K\bar K$
bound states? As for the $\sigma$, authors do not even agree on
its existence as a fundamental hadron, although the number of
supporters is growing rapidly.

In Fig. 1 we have plotted with filled circles the results of 22
different analyses on the \sig\ pole position, which are included
in the  2000 edition of the Review of Particle Physics
\cite{pdg2000} under the entry $f_0(400-1200)$ or \sig. Most of
these find a \sig\ pole position near 500-i250 MeV.

Also, at  a recent meeting \cite{kyoto} devoted to the $\sigma$,
many groups reported preliminary analyzes, which find the \sig\
resonance parameters in the same region. These are plotted as
triangles in Fig. 1. It was not possible to here distinguish
between Breit-Wigner parameters and pole positions, which of
course can differ by several 100 MeV for the same data.
 It must also be noted that many of the triangles in Fig. 1 rely on the same
raw data and come from preliminary analyzes not yet published.

We also included in Fig. 1 (with a star) the \sig\ parameters
obtained from the recent E791 Experiment at Fermilab \cite{E791},
where 46\% of the $D^+\to3\pi$ Dalitz plot is $\sigma\pi$. The
open circle in the same figure represents the \sig\ parameters
extracted from the CLEO analysis of $\tau\to\sigma\pi\nu\to
3\pi\nu$ \cite{CLEO}.

\begin{figure}
% \epsfxsize=13 cm
% \epsfysize=13 cm
%\centerline{\null\hskip 1cm \epsffile{sigmapoles.eps}}
%\centerline{\null\hskip 1cm \epsfig{sigmapoles.eps}}
\begin{center}
\includegraphics*[width=13cm]{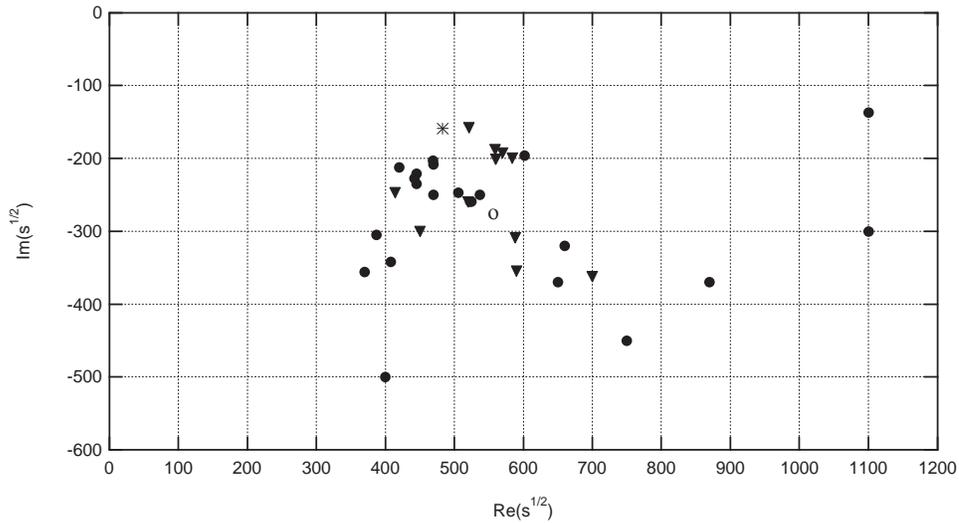}
\end{center}
%\vskip 1.5 cm
\caption{ The pole positions of the \sig\ resonance, as listed by
the PDG \cite{pdg2000} under $f_0(400-1200)$ or
 \sig\ (filled circles), plotted in the complex energy plane (in units of MeV).
The triangles represent the mass and width parameters (plotted as
$m-i\Gamma /2$), which were reported at this meeting. We could not
here distinguish between pole and Breit-Wigner parameters. The
star is the $m-i\Gamma /2$ point obtained from the recent E791
experiment \cite{E791} on $D\to\sigma \pi\to 3\pi$ ($m_\sigma=478$
MeV, $\Gamma_\sigma = 324$ MeV) while the open circle is that
obtained by the CLEO analysis of $\tau\to\sigma\pi\nu\to 3\pi\nu$
\cite{CLEO}.}
\end{figure}

\section{The NJL and the Linear sigma Model}

The NJL model is an effective theory which is believed to be
related to QCD at low energies, when one has integrated out the
gluon fields. It involves a linear realization of chiral symmetry.
After bosonization of the NJL model one finds essentially the
Linear sigma Model (\lsm ) as an approximate effective theory for
the scalar and pseudoscalar meson sector.

About  30 years ago Schechter and Ueda \cite{u3u3}  wrote down the
$U3\times U3$ \lsm\ for the meson sector involving a scalar and a
pseudoscalar nonet. This (renormalizable) theory has only 6
parameters, out of which 5 can be fixed by the pseudoscalar masses
and decay constants ($m_\pi,\ m_K, \ m_{\eta^\prime},\ f_\pi, \
f_K$). The sixth parameter for the OZI rule violating 4-point
coupling must be small. One can then predict, with no free
parameters, the tree level scalar masses \cite{lsm}, which turn
out to be not far from the lightest experimental masses, although
the two quantities are \underline{not} exactly the same thing but
can differ for the same model and data by over 100 MeV.

The important thing is that the scalar masses are predicted to be
near the lightest experimentally seen scalar masses, and not in
the 1500 MeV region where many authors want to put the lightest
$q\bar q$ scalars. The \sig\ is predicted \cite{lsm} at 620 MeV
with a very large width ($\approx 600$ MeV), which well agrees
with Fig. 1. The $a_0(980)$ is predicted at 1128 MeV,  the
$f_0(980)$ at 1190 MeV, and the $K^*_0(1430)$ at 1120 MeV, which
is surprisingly good considering that loop  effects are large.

\section{ Understanding the S-waves within a unitarized quark model (UQM)}
In Figs.~2-4 we show the obtained fits to the $K\pi$, $\pi\pi$
S-waves and to the  \az\ resonance peak in $\pi\eta$. The Partial
Wave Amplitude (PWA) in the case of one \qq\ resonance, such as
the  $a_0(980)$, can be written as:
\begin{equation} A(s)=-\frac{Im\Pi_{\pi\eta}(s)}{[m_0^2+Re\Pi
(s)-s +iIm\Pi (s)]},
\end{equation}
where: \bea \label{PWA} Im\Pi (s)&=&\sum_i Im\Pi_i(s) \label{impi}
=-\sum_i \gamma_i^2(s-s_{A,i})\frac{k_i}{\sqrt s}e^{-k_i^2/k_0^2}
           \theta(s-s_{th,i})\ ,\nonumber \\
Re\Pi (s)&=&\frac 1\pi{\rm P.V.}\int^\infty_{s_{th,1}} \frac{Im\Pi
(s)}{s'-s} ds' \ . \nonumber \eea Here the coupling constants
$\gamma_i$ are related by flavour symmetry and OZI rule, such that
there is only one over all parameter $\gamma$. The $s_{A,i}$ are
the positions of the Adler zeroes, which  are  near $s=0$. Eq. (1)
can be looked upon as a more general Breit-Wigner form, where the
mass parameter is replaced by an $s$-dependent function, ``the
running mass" $m_0^2 +  Re\Pi (s)$.

In the flavourless channels the situation is a little more
complicated than in Eq. (1) since one has both \uu\ and \ss\
states, requiring a two dimensional mass matrix (see
Ref.~\cite{NAT2}).
 Note that the sum runs over all light
PP thresholds, which means three for the \az : $\pi\eta ,\ K\bar
K,\pi\eta'$ and three for the \Kz : $ K\pi ,\ K\eta ,\ K\eta' $,
while for the $f_0$'s there are five channels: $\pi\pi ,K\bar K ,\
\eta\eta ,\ \eta\eta' ,\ \eta'\eta'$.

In Fig. 5 we show, as an example, the running mass,
$m_0^2+Re\Pi(s)$,
 and the width-like function,
$-Im\Pi(s)$, for the I=1 channel. The crossing point of the
running mass with $s$ gives the $90^\circ$  mass of the \az.
 The magnitude of the \KK\ component in the \az\ is determined by
$-\frac d{ds}Re\Pi(s)$, which is large in the resonance region
just below the \KK\ threshold. These functions fix the PWA of Eq.
(1) and Fig.~3. In Fig. 6 the running mass and width-like function
for the strange channel are shown. These fix the shape of the
$K\pi$ phase shift and absorption parameters in Fig. 1. As can be
seen from Figs.~1-3, the model gives a good description of the
relevant data.

In Ref.~\cite{NAT2} the \sig\ was missed because only poles
nearest to  the physical region were looked for, and the
possibility of the resonance doubling phenomenon, discussed below,
was overlooked. Only a little later we realized with Roos
\cite{NAT1} that two resonances (\fz\ and $ f_0(1370)$) can emerge
although only one $s\bar s$ bare state is put in.

In fact, it was pointed out by Morgan and Pennington~\cite{morgan}
that for each \qq\ state there are, in general, apart from the
nearest pole, also  image poles, usually located far from the
physical region. As explained in more detail in Ref.~\cite{NAT1},
some of these can (for a large enough coupling and sufficiently
heavy threshold) come so close to the physical region that they
make new resonances. And, in fact, there are more than four
physical poles  with different isospin, in the output spectrum of
the UQM model, although only four bare states are put in!. The
\fz\ and the \fzz\ of the model thus turn out to be  two
manifestations of the same \ss\ state (see \cite{NAT2} and Fig. 7
for details). There can be two crossings with the running mass
$m_0^2+Re\Pi(s)$, one near the threshold and another at higher
mass, and each one is related to a different pole at the second
sheet (or, if the coupling is strong enough, the lower one could
even become a bound state pole, below the threshold, on the first
sheet).

 Similarly the \az\ and the  \azz\ could be two manifestations of the $u\bar d$ state.
Only after  realizing that this resonance doubling is important we
looked deeper into the second sheet and found the light and broad
\sig\ \cite{NAT1}.

Another important effect that the model can explain is the large
mass difference between the $a_0$ and $K^*_0$. Because of this
large splitting many authors argue that the $a_0(980)$ and
$f_0(980)$ are not \qq\ states, since in addition to being very
close to the \KK\ threshold, they are much lighter than the first
strange scalar, the  $K^*_0(1430)$. Naively one expects a mass
difference between the strange and nonstrange meson to be of the
order of the strange-nonstrange quark mass difference, or a little
over 100 MeV. This is also one of the reasons why some authors
want to have a lighter strange meson, the $\kappa$, near 800 MeV.
Cherry and Pennington \cite{cherry} recently have strongly argued
against its existence.

Figs. 5 and 6 explain why  one can easily understand this large
mass splitting as a secondary effect of the large pseudoscalar
mass splittings, and because of the large mass shifts coming from
the loop diagrams involving the PP thresholds. If one puts Figs. 4
and 5 on top of each other one  sees that  the 3 thresholds
$\pi\eta,\ K\bar K,\ \pi\eta$ all lie relatively close to the
$a_0(980)$, and all 3 contribute to a large mass shift. On the
other hand, for the $K^*_0(1430)$, the $SU3_f$ related thresholds
($K\pi,\ K\eta'$) lie far apart from the $K^*_0$, while the
$K\eta$ nearly decouples because of the physical value of the
pseudoscalar mixing angle.

\section{$D\to\sigma\pi\to 3\pi$}

The recent experiments studying charm decay to light hadrons are
opening up a new experimental window for understanding light meson
spectroscopy and especially the controversial scalar mesons, which
are copiously produced in these decays.

In particular we refer to the E791 study of the $D\to 3\pi$ decay
\cite{E791} where it is shown how adding an intermediate scalar
resonance with floating mass and width in the Monte Carlo program
simulating the  Dalitz plot densities, allows for an excellent fit
to data provided the mass and the width of this scalar resonance
are $m_\sigma\simeq 478$ MeV and $\Gamma_\sigma\simeq 324$ MeV.
This resonance is a very good candidate for the $\sigma$. To check
this hypothesis we adopt the E791 experimental values for its mass
and width and using a Constituent Quark Meson Model (CQM) for
heavy-light meson decays \cite{rass} we compute the
$D\to\sigma\pi$ non-leptonic process via {\it factorization}
\cite{WSB} and taking the coupling of the $\sigma$ to the light
quarks from the Linear sigma Model \cite{volkoff}. In such a way
one is directly assuming that the scalar state needed in the E791
analysis could be the quantum of the $\sigma$ field of the Linear
sigma Model. According to the CQM model and to factorization, the
amplitude describing the $D\to\sigma\pi$ decay can be written as a
product of the semileptonic amplitude $\langle
\sigma|A^\mu_{(\bar{d}c)}(q)|D^+\rangle $, where $A^\mu$ is the
axial quark current, and $\langle \pi|A_{\mu(\bar{u}d)}(q)|{\rm
VAC}\rangle$. The former is parameterized by two form factors,
$F_1(q^2)$ and $F_0(q^2)$, connected by the condition
$F_1(0)=F_0(0)$, while the latter is governed by the pion decay
constant $f_\pi$. As far as the product of the two above mentioned
amplitudes is concerned, only the form factor $F_0(q^2)$ comes
into the expression of the $D\to\sigma\pi$ amplitude. Moreover we
need to estimate it at $q^2\simeq m_\pi^2$, that is the physically
realized kinematical situation. CQM offers the possibility to
compute this form factor through two quark-meson 1-loop diagrams
that we call the {\it direct} and the {\it polar} contributions to
$F_0(q^2)$. These quark-meson loops are possible since in the CQM
one has effective vertices (heavy quark)-(heavy meson)-(light
quark) that allow to {\it compute} spectator-like diagrams in
which usually the external lines represent incoming or outgoing
heavy mesons while the internal lines are the constituent light
quark and heavy quark propagators.

In Figs. 8 and 9 we show respectively the {\it direct} and the
{\it polar} diagrams for the semileptonic amplitude $D\to\sigma$,
the former being characterized by the axial current directly
attached to the constituent quark loop, the latter involving an
intermediate $D(1^+)$ or $D(0^-)$ state. These two diagrams are
computed with an analogous technique and one finally obtains a
determination of the direct and polar form factors $F_0^{\rm dir,
pol}(q^2)$. The extrapolation to $q^2\simeq m_\pi^2\simeq 0$ is
safe for  the direct form factor while is not perfectly under
control for the polar form factor since the latter is more
reliable at the pole $q^2\simeq m_P^2$, $m_P$ being the mass of
the intermediate state in Fig. 9. We take into account the
uncertainty introduced by this extrapolation procedure and
signaled by the fact that we find $F_0^{\rm pol}(0)\neq F_1^{\rm
pol}(0)$ (computing $F_0$ from the polar diagram with $0^-$
intermediate polar state and $F_1$ from that with intermediate
$1^+$ state). Our estimate for $F_0(0)=F_0^{\rm pol}(0)+F_0^{\rm
dir}(0)=0.59\pm 0.09$ is in reasonable agreement with an estimate
of $F_0(m_\pi^2)=0.79\pm 0.15$ carried out in \cite{dib} using the
E791 data analysis and a Breit-Wigner like approximation for the
$\sigma$.

This computation indicates that the scalar resonance described in
the E791 paper can be consistently understood as the $\sigma$ of
the Linear sigma Model. Of course a calculation such as the  one
here described calls for alternative calculations and/or
explanations of the E791 data for a valuable and useful comparison
of point of views on the $\sigma$ nature.

\section{Concluding remarks}
An often raised question is: Why are the mass shifts required by
unitarity so much more important for the scalars than, say, for
the vector  mesons? The answer is very simple, and there are two
main reasons apart from chiral symmetry constraints:
\begin{itemize}
\item
The scalar coupling to two pseudoscalars is very much larger than
the corresponding coupling for the vectors, both experimentally
and theoretically (e.g., spin counting gives  3 for the ratio of
the two squared couplings).
\item For the scalars the thresholds are S-waves,
giving nonlinear square root cusps in the $\Pi(s)$ function,
whereas for the vectors the thresholds are P-waves, giving a
smooth $k^3$ angular momentum and phase space factor.
\end{itemize}
\vskip .2cm

One could argue that the two states \fz\ and \az\ are a kind of
$K\bar K$ bound states (c.f. Ref.~\cite{wein}), since these have a
large component of virtual $K\bar K$ in their wave functions.
However, the dynamics of these states is quite different from that
of normal two-hadron bound states. If one wants to consider them
as \KK\ bound states, it is the $K\bar K \to s\bar s \to K\bar K$
interaction which  creates their binding energy, not the hyperfine
interaction as in Ref.~\cite{wein}. Thus, although they may spend
most of their time as $K\bar K$, they owe their existence to the
\ss\ state. Therefore, it is more natural to consider the \fz\ and
\fzz\ as two manifestations of the same \ss\ state.

The wave function of the $a_0(980)$ (and $f_0(980)$) can be
pictured as a relatively small core of \qq\ of typical \qq\ meson
size (0.6fm), which is  surrounded by a much larger standing
S-wave of virtual \KK. This picture also gives a physical
explanation of the narrow width: in order to decay to $\pi\eta$,
the \KK\ component must first virtually annihilate near the origin
to \qq. Then the \qq\ can decay to $\pi\eta$ as an OZI allowed
decay.

\section*{Acknowledgements}
NAT and ADP acknowledge support  from EU-TMR programme, contract
CT98-0169. The authors are also grateful to A. Deandrea, R. Gatto
and G. Nardulli for useful discussions.

\twocolumn[\hsize\textwidth\columnwidth\hsize\csname
@twocolumnfalse\endcsname]
%\null
\begin{figure}[t!]
\includegraphics[width=15pc]{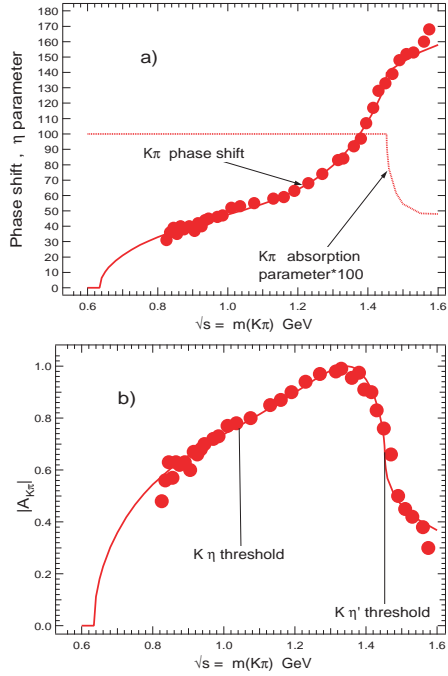}
\caption{  \footnotesize (a) The $K\pi$ S-wave phase shift and (b)
the magnitude of the $K\pi$ PWA compared with the model
predictions, which fix 4 ($\gamma$, $m_0+m_s$, $k_0$ and
$s_{A,K\pi}$) of the 6 parameters. The parameters here fixed are
then used in the analysis of Fig. 3. For more details see
\cite{NAT1,NAT2}.}
\end{figure}
%===============================================================================
\begin{figure}[t!]
\vspace{0cm}
\includegraphics[width=15pc]{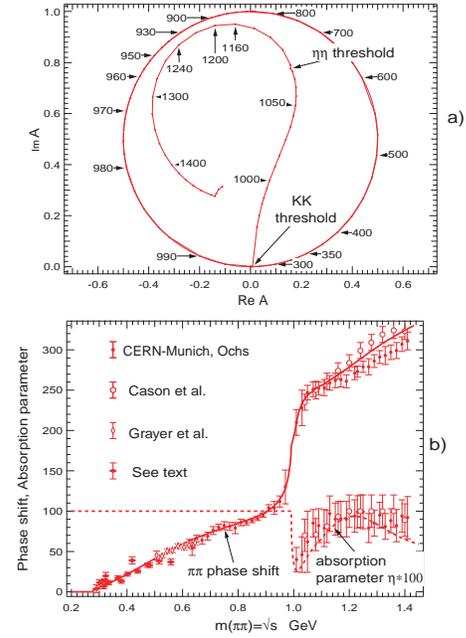}
%\epsfxsize=5.3 cm \epsfysize=7. cm \hskip 0.8cm
%\epsffile{fig2bruss.eps}
%\psfig{figure=fig2bruss.epsf,height=3.5in}
\caption{  \footnotesize (a) The $\pi\pi$ Argand diagram and (b)
phase shift predictions are compared with data.
%Note that most of the
%parameters were fixed by the data in Fig. 1. For more details see
%Ref.~$^{1,2}$.
}
\end{figure}
%================================================================================
\vspace{1cm}
\begin{figure}[t!]
\includegraphics[width=15pc]{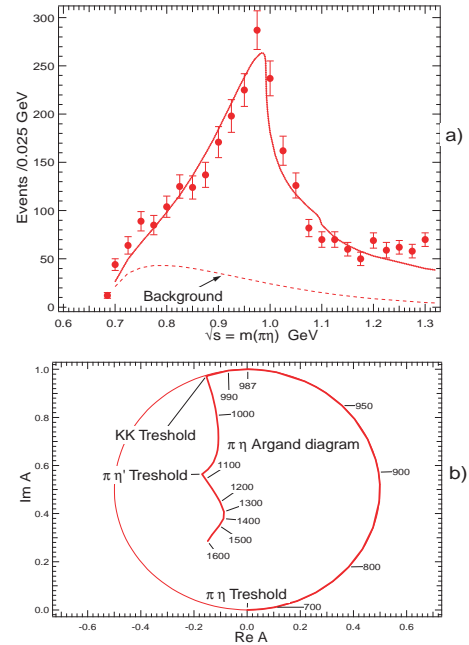}
%\epsfxsize=5. cm \epsfysize=6 cm \hskip .8 cm
%\epsffile{fig3bruss.eps}
%\psfig{figure=fig3bruss.epsf,height=3.3in}
\caption{ \footnotesize (a) The $a_0(980)$ peak compared with
model prediction and (b) the predicted $\pi\eta$ Argand diagram.}
\end{figure}

%=================================================================================
\begin{figure}
\includegraphics[width=15pc]{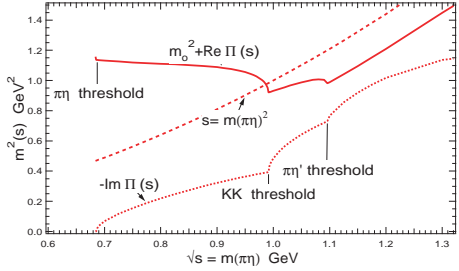}
%\epsfxsize=4.7 cm \epsfysize=3 cm \hskip 0.9cm
%\epsffile{fig4bruss.eps}
%\psfig{figure=fig4bruss.epsf}
\caption{ \footnotesize The running mass $m_0+ Re\Pi(s)$ and $Im
\Pi (s)$ of the $a_0(980)$. The strongly dropping running mass at
the $a_0(980)$ position, below the $K\bar K$ threshold,
contributes to the narrow shape of the peak in Fig. 3a.}
\end{figure}
%=================================================================================
\begin{figure}
\includegraphics[width=15pc]{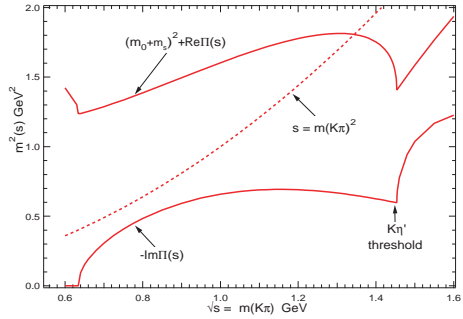}
\caption{ \footnotesize The running mass and width-like function
$-Im \Pi(s)$ for the $K^*_0(1430)$. The crossing of $s$ with the
running mass gives the 90$^\circ$ phase shift mass, which roughly
corresponds to a naive Breit-Wigner mass, where the running mass
is put constant. }
\end{figure}
%===============================================================================
\begin{figure}[t!]
\includegraphics[width=15pc]{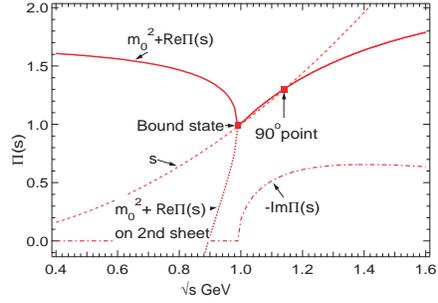}
\caption{\label{fig:fig1} \footnotesize
         Although the model has only one bare $s\bar s $
resonance, when unitarized it can give rise to two crossings with
the running mass in the $s\bar s - K\bar K$ channels. This means
the $s\bar s$ state can manifest itself in two physical
resonances, one at threshold and one near 1200 MeV. }
\end{figure}
%===============================================================================
\begin{figure}[t!]
\includegraphics[width=15pc]{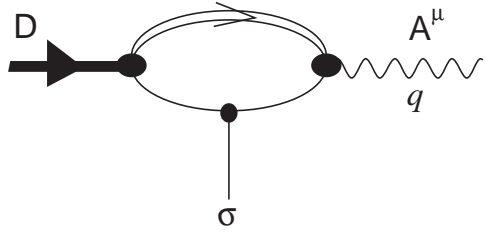}
\caption{\label{fig:fig1} \footnotesize
          Diagram for the {\it direct} contribution to the $D\to\sigma$ semileptonic
          amplitude. The axial current is directly attached to the quark loop. }
\end{figure}
%===============================================================================
%===============================================================================
\begin{figure}[t!]
\includegraphics[width=15pc]{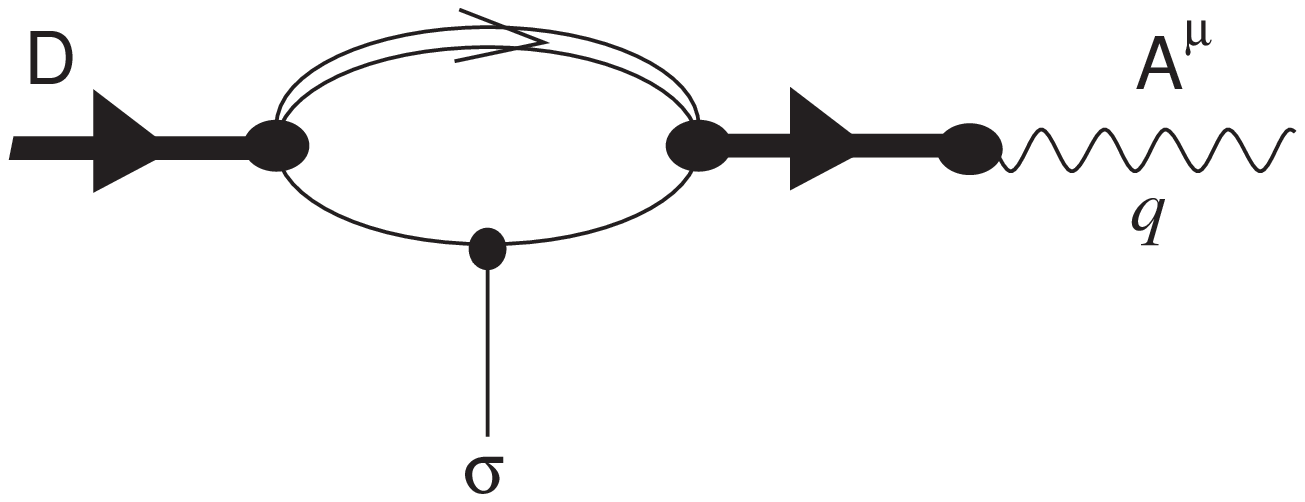}
\caption{\label{fig:fig2} \footnotesize
         The {\it polar} contribution to $F_0$, if a $0^-$ intermediate
         state is considered, and to $F_1$, with a $1^+$ intermediate state. The
         $D(1^+)$ state is described in the PDG \cite{pdg2000}.}
\end{figure}
%===============================================================================

\begin{thebibliography}{99}
\bibitem{NAT1} N.A. T\"ornqvist and M. Roos, Phys. Rev. Lett. {\bf 76}, 1575
(1996).
\bibitem{NAT2}
N.A. T\"ornqvist, Z. Phys {\bf C68}, 647 (1995);
 N. A. T\"ornqvist, Eur. J. Phys. {\bf C11}, 359 (1999).
\bibitem{gatto} R. Gatto, G. Nardulli, A.D. Polosa and N.A. T\"ornqvist,
hep-ph/0007207, in press in Phys. Lett. {\bf B}.
%\bibitem{lsm} J. Schwinger, Ann. Phys. {\bf 2} 407 (1957); M. Gell-Mann, M.
%Levy, Nuovo Cim. {\bf 16} 705 (1960); B.W. Lee Nucl. Phys. {\bf B 9} (1969); J. Schechter and Y. Ueda, Phys.Rev. {\bf D 3} (1971) 2874.
%\bibitem{pdg1} A. Rosenfeld et al. Rev. Mod. Phys. {\bf 36} (1964) 977.
%\bibitem{samois} N.P. Samois et al. Phys. Rev. Lett. {\bf 9} (1962) 139.
\bibitem{pdg2000} D.E. Groom et al. Eur. J. Phys. {\bf C 15}, 1 (2000).
\bibitem{kyoto} Conference: ``Possible existence of the light $\sigma$
resonance and its implications to hadron physics", Kyoto, Japan
11-14th June 2000, KEK-proceedings/2000-4; See also N.A.
T\"ornqvist, ``Summary of the conference", hep-ph/0008135.
\bibitem{E791} E. M. Aitala et al. (E791 collaboration),
{\it Experimental evidence for a light and broad scalar resonance
in $D^+\to 3\pi$}, hep-ex/0007028.
\bibitem{CLEO} D.M. Asner {\it et al.} (CLEO collaboration), Phys.
Rev. {\bf D61}, 0120002 (1999).
\bibitem{u3u3} J. Schechter and Y. Ueda, Phys. Rev. {\bf D3}, 2874 (1971).
\bibitem{lsm} N.A. T\"ornqvist, Eur. J. Phys. {\bf C11}, 559 (1999);
M. Napsusciale, hep-ph/9803396; see also G. Parisi and M. Testa
Nuov. Cim. {\bf LXVII}, 13 (1969).
\bibitem{cherry} S. N. Cherry and M. Pennington, hep-ph/0005208.
\bibitem{morgan} D. Morgan and D. Pennington, Phys. Rev. {\bf D48}, 1185
(1993); {\it ibid.} {\bf D48}, 5422 (1993).
\bibitem{wein}  J. Weinstein and N. Isgur, Phys. Rev. Lett. {\bf 48}, 659
(1982);  Phys. Rev. Lett. {\bf 27}, 588 (1983).
\bibitem{rass} A.D. Polosa, {\it The CQM model}, hep-ph/0004183
and references therein.
\bibitem{WSB} M. Bauer, B. Stech and M. Wirbel, Z. Phys. {\bf
C16}, 205 (1983).
\bibitem{volkoff} D. Ebert and M.K. Volkov, Z. Phys. {\bf C16}, 205
(1983).
\bibitem{dib} C. Dib and R. Rosenfeld, {\it Estimating $\sigma$
meson couplings from $D\to 3\pi$ decays}, hep-ph/0006145.
\end{thebibliography}
\end{document}